# Preparation of large Cu$_3$Sn single crystal by Czochralski method

*Minsik Kong[1, †], Sang-Eon Park[2,3, †], Hye Jung Kim,[1,†] Sehwan Song[1], Dong-Choon Ryu[4,5], Baekjune Kang[6], Changhee Sohn[6], Hyun Jung Kim[2], Youngwook Kim[7], Sangmoon Yoon[8], Ara Go[9], Hyoungjeen Jeen[1], Sungkyun Park[1], Se-Young Jeong[10,11], Chang-Jong Kang[4,\*], and Jong Mok Ok[1,\*]*

[1]Department of Physics, Pusan National University, Busan 46241, Republic of Korea

[2]Quantum Matter Core-Facility, Pusan National University, Busan 46241, Republic of Korea.

[3]Crystal Bank Research Institute, Pusan National University, Busan 46241, Republic of Korea

[4]Department of Physics, Chungnam National University, Daejeon 34134, Republic of Korea

[5]Department of Physics, Kunsan National University, Gunsan 54150, Republic of Korea

[6]Department of Physics, Ulsan National Institute of Science and Technology; Ulsan, Republic of Korea.

[7]Department of Physics and Chemistry, DGIST, 42988 Daegu, Republic of Korea

[8]Department of Physics, Gachon University, Seongnam13120, Republic of Korea

[9]Department of Physics, Chonnam National University, Gwangju, 61186, Republic of Korea

[10]Department of Cogno-Mechatronics Engineering, Pusan National University, Busan 46241, Republic of Korea

[11]Department of Optics and Mechatronics Engineering, Pusan National University, Busan 46241, Republic of Korea

[†]These authors contributed equally to this work.

[\*]Corresponding author: Email: cjkang87@cnu.ac.kr, okjongmok@pusan.ac.kr

**KEYWORDS.** Single crystal, Czochralski, Intermetallic, Cu$_3$Sn

**ABSTRACT.**

$Cu_3Sn$ was recently predicted to host topological Dirac fermions, but related research is still in its infancy. The growth of large and high-quality $Cu_3Sn$ single crystals is, therefore, highly desired to investigate the possible topological properties. In this work, we report the single crystal growth of $Cu_3Sn$ by Czochralski (CZ) method. Crystal structure, chemical composition, and transport properties of $Cu_3Sn$ single crystals were analyzed to verify the crystal quality. Notably, compared to the mm-sized crystals from a molten Sn-flux, the cm-sized crystals obtained by the CZ method are free from contamination from flux materials, paving the way for the follow-up works.

**INTRODUCTION.**

Over the past decade, topological materials have attracted significant attention, because their ultrahigh mobility and magnetoresistance may lead to the next generation of device applications [1-2]. In topological insulators, the current flow only along their surface while bulk exhibits semiconducting or insulating behavior with a band gap. In this context, topological semimetals are more suitable for the high mobility device application due to their bulk Dirac point originating from the crossing of conduction and valence bands. Thus, searching the applicable topological semimetal is currently one of the most interesting issues. Furthermore, a more important point in terms of finding such materials is that it is typically composed of inexpensive and nontoxic elements and can be grown to large-sized single crystals.

In this point of view, Cu-Sn alloys, so-called bronze, can be a promising candidate. Cu-Sn alloy has been used for the longest time in human history, and the discovery of this alloy allowed humankind opened the Bronze Age. Cu-Sn alloys have been studied a lot, especially for high-temperature applications, owing to their excellent thermodynamic and kinetic stability [3]. On the other hand, their topological properties do not attract lots of interest due to lack of understanding, however they become major subjects of great interest recently [4]. Among the Cu-Sn alloys, a Cu$_3$Sn phase can crystallize as three different structures including the orthorhombic [5-6], hexagonal [7], and cubic structures [8], which makes this system particularly interesting. Based on the recent research, all the Cu$_3$Sn phases can be categorized into topological semimetals [9]. More recently, C. Liu *et al*. reported the observation of clear de Haas-van Alphen oscillations in orthorhombic Cu$_3$Sn single crystals grown by the self-flux method suggesting it exhibits a nontrivial band topology. Interestingly, the mobility of huge hole carrier ($n_h$ ~ 2.68×10$^{22}$ cm$^{-3}$) in the Cu$_3$Sn single crystal is estimated to be $\mu$ ~ 490 cm$^2$V$^{-1}$s$^{-1}$, which is an order of magnitude higher than that of electron carriers in the topologically trivial Cu single crystal $\mu$ ~ 35 cm$^2$V$^{-1}$s$^{-1}$ [10]. This study additionally suggested the possibility of a bronze-based topological semimetal with highly mobile carriers.

Herein, we report the growth of a large-sized intermetallic Cu$_3$Sn single crystal by the Czochralski (CZ) method. The CZ method is one of the powerful tools for the growth of the large-scale ultrapure single crystal and has contributed to the growth of a lot of materials, such as metal oxides [11], heavy fermion systems [12], and chiral compounds [13]. Compared to the crystals grown by the flux method, whose notable downside is that the flux can be trapped inside the crystal or remain on the edges or surface, the crystals grown by CZ method are free of contamination. The contamination from the flux can deteriorate the physical properties of the crystal, hindering the understanding of intrinsic behavior. The Cu$_3$Sn single crystal grown by the CZ method is around several centimeters, which is much larger than the self-flux grown crystals [5]. Based on the X-ray diffraction (XRD) and scanning electron microscopy with energy-dispersive spectroscopy (EDS), the crystal structure of the obtained single crystal was confirmed to be orthorhombic $Cu_3Sn$ with space group *Pmmn*.

**EXPERIMENTS**

Intermetallic single crystals of $Cu_3Sn$ were grown by the Czochralski method using a single crystal growth system installed in Quantum Matter Core-Facility of Pusan National University. Cu-Sn compounds were weighed according to the stoichiometric ratio of Cu: Sn and 3: 1.1 at %. The excess Sn is used to compensate for the volatilization of Sn during the single crystal growth at high temperatures. The raw materials for growing $Cu_3Sn$ were heated in a 40 kHz RF induced coil system. A graphite crucible whose diameter and height are 45 mm and 50 mm was used for the crystal growth. Growths were carried out several times by re-melting synthesized materials and introducing seeds. Seeds were cut first from polycrystalline materials prepared in the first synthesis run and later from grown crystals. The typical size of seeds used in growth was 3×3×30 mm. The pulling rate is 7mm/h, and the seed rotation rate is 12 rpm. The growth was conducted in an Ar atmosphere to minimize the evaporation of Sn. After the growth, the single crystal was cooled down to room temperature with a ramp rate of 100-150 ℃/h.

The crystal structure and chemical composition were characterized by X-ray diffraction (XRD) and scanning electron microscopy with energy-dispersive spectroscopy (SEM-EDS). XRD patterns were measured for single crystals and crushed powders using a Philips X'pert3 and Bruker D8 Discover X-ray diffractometer with Cu-K$\alpha$1 radiation ($\lambda$ = 0.15406 nm) in the range of 20-80°. The electronic transport measurements were performed with a 9T cryostat. The electrical contacts were made with silver paste/epoxy in a 4-probe configuration, and the field-dependent data were symmetrized. First principles calculations are performed using the Vienna

Ab into Simulation Package (VASP). In this calculation, we used the projector augmented wave (PAW) method and the exchange-correlation functional obtained from the general gradient approximation (GGA). The fully structural optimization was done with 19✕17✕15 k-mesh and cutoff energy 600 eV.

**RESULTS & DISCUSSIONS**.

A series of runs were carried out to obtain single crystal $Cu_3Sn$ as described in Fig. 1. The first run was the simple thermal process to obtain a polycrystalline seed. The seeds used for the second and third growths were obtained from the first and second growths, respectively. We have tried growth a total 5 times, but as the growth was repeated, the ratio of Cu to Sn was changed due to the evaporation of Sn. The best samples were obtained from the third growth. Figure 2(a) shows an optical image of the best $Cu_3Sn$ single crystal grown by the CZ method. The size of the crystal is around $1\times1\times3$ cm$^3$. Figures 2(b) and 2(c) show the XRD pattern of $Cu_3Sn$ obtained from the second and third runs. $Cu_3Sn$ crystal obtained from the second run is found to have multiple grain domains, as shown in Fig. 2(b). We noted that all XRD peaks found in the crystal from the second run are well matched with orthorhombic $Cu_3Sn$. The grains are mainly characterized by (102) and (321) orientations among the multiple domains.

To obtain a better quality of $Cu_3Sn$ crystal, the third run was carried out by using the seed with multiple grains obtained from the second run. Most grains are eliminated, while only grains oriented in the (102) direction are left, as confirmed by XRD displayed in Fig. 2(c). We note that the crystal grown in the (102) direction is unusual. This observation indicates that the

(012) direction is the possibly preferred growth direction of Cu$_3$Sn. The (012) direction is the most intense peak also measured in powder XRD, shown later.

The structural characterization of obtained Cu$_3$Sn single crystals is carried out using powder XRD. The XRD pattern of crushed crystals is analyzed using Rietveld refinement. Figure 2(d) shows the XRD data, Rietveld fit profile, Bragg positions, and difference in experimental and refined curve intensities. The power XRD shows sharp peaks only corresponding to orthorhombic Cu$_3$Sn, suggesting that the sample is a single phase with good crystalline quality. The powder XRD data is well refined by the *Pmmn* (No. 59) space group with lattice parameters of $a = 4.34$ Å, $b = 4.86$ Å, and $c = 5.48$ Å, which are very close to the reported lattice parameter of Cu$_3$Sn for the orthorhombic *Pmmn* phase. More detailed information obtained from the refinement is shown in Table 1.

The chemical composition of the Cu$_3$Sn single crystal was analyzed by SEM-EDS. Figures 3(a) and 3(b) shows a typical EDS spectrum and SEM image of a polished surface of the Cu$_3$Sn single crystal. Cu and Sn atoms are detected from the EDS spectrum. The composition is determined by averaging over 5 points of the same specimen on the polished surface and for several crystals. The average atomic ratio of the Cu$_3$Sn single crystal is Cu: Sn = 3.05:1. By considering the typical error of SEM/EDX (typical error $\pm$ 5%), these results suggest that nearly stoichiometric crystals can be achieved by the CZ method. Furthermore, the result suggests that Cu$_3$Sn crystals obtained from the CZ method are free from Sn contamination. The chemical homogeneity of Cu$_3$Sn single crystal was further confirmed by elemental mapping, as displayed in Fig. 3(c). The mapping was conducted on the sample surface with a dimension of about $10 \times 8$ μm$^2$. The results revealed a uniform distribution of Cu and Sn, suggesting that the surface of the Cu$_3$Sn crystal is compositionally homogenous, at least within the spatial resolution of SEM-EDS measurement.

In order to investigate the transport property of Cu$_3$Sn crystals, measurements of the electrical resistivity and magnetoresistance were conducted. Since the crystals obtained by the CZ method are large (1×1×3 cm$^3$), the crystal was cut into a rectangular shape by electric discharge machining (EDM). The smallest dimension of the sample that can be prepared by EDM is ~3×3×10 mm$^3$; however, the resistance of the sample was lower than the limit of the measurement device due to its high degree of metallicity of 0.1-0.3 μΩcm at low temperature [5]. As a last resort to allow the transport measurements, the crystals were crushed into small pieces with a size of ~0.3×0.3×1 mm$^3$. However, it was difficult to obtain accurate transport data of the tiny Cu$_3$Sn pieces due to the unclear orientation and dimensions of the sample. The transport properties of Cu$_3$Sn are highly dependent on the crystal orientation, as previously reported [5]. Although the crystal orientation of the measured piece is not well defined, intriguing transport properties similar to those reported previously have been observed [5].

Figure 4(a) displays the temperature dependence of resistance $R$(T) of the small Cu$_3$Sn piece. Cu$_3$Sn shows a metallic behavior and a power-law dependence at low temperatures. The solid red line is the fitted curve using $R = AT^\alpha$. The estimated value of α from the best fit is 2.6, which agrees with the previous report [5]. Figure 4(b) shows magnetic field dependence of magnetoresistance (MR) at several temperatures. MR of Cu$_3$Sn sample shows quasi-linear field dependence without saturation, which is again consistent with the previous result [5]. However, at high temperatures, the non-saturating quasi-linear MR gradually changes to conventional quadratic MR, indicating that the unusual MR of Cu$_3$Sn is related to its low-energy electronic structure. Such behavior is usually an indication of non-trivial low-energy electronic structures, for example perfectly compensated electron/hole Fermi surface [14-15], strong inhomogeneity-induced charge/mobility fluctuation [16], and topological Dirac/Weyl semimetals [17-19].

Among the possibilities, the quasi-linear MR of $Cu_3Sn$ is likely due to its topological electronic band structure. In order to investigate the topological property of $Cu_3Sn$, we examine the parity of each energy eigenstate at 8 time-reversal invariant momenta (TRIM) points displayed in the inset of Fig. 5. Note that band degeneracy at 8 TRIM points except $\Gamma$ and Y is 4 (including the spin degeneracy), so that parities at these TRIM points do not affect the topological $Z_2$ index. Only at two TRIM points, $\Gamma$ and Y, energy eigenstates are doubly degenerate (time-reversal pair) and their parities determine whether the $Z_2$ index is trivial or not. Figure 5 presents the band dispersion of $Cu_3Sn$ with the inclusion of the spin-orbit coupling. It also shows band representations with parity eigenvalues at $\Gamma$ and Y around the Fermi level. Table 2 shows the products of parity eigenvalues of the occupied states for high symmetry k-points in the bulk Brillouin zone. It clearly demonstrates that band/parity inversion occurs at $\Gamma$, thereby providing the nontrivial $Z_2$ index. These results suggest that $Cu_3Sn$ is a non-trivial $Z_2$ topological semimetal.

**CONCLUSION.**

In summary, high-quality $Cu_3Sn$ single crystals in cm-size are successfully synthesized by the CZ method. The structural and chemical composition measurements confirm the high quality and homogeneity of $Cu_3Sn$ crystals. Although the electronic transport properties were not precisely measured due to the difficulties in sample preparation, non-saturating quasi-liner MR, which supports the possibility of Dirac band, and is in good agreement with previous reports, was successfully observed. However, more efforts to prepare samples with well-defined geometries and crystal orientations are needed to gain better insight into the topological properties of Cu$_3$Sn. The large-sized single crystals free from contamination of Sn flux will enable the various measurements, including neutron scattering, magnetization, specific heat, and optical conductivity, thereby promoting more in-depth studies of the topological properties of Cu$_3$Sn.

**Acknowledgment**

This research was supported by PNU-RENovation (2021-2022).

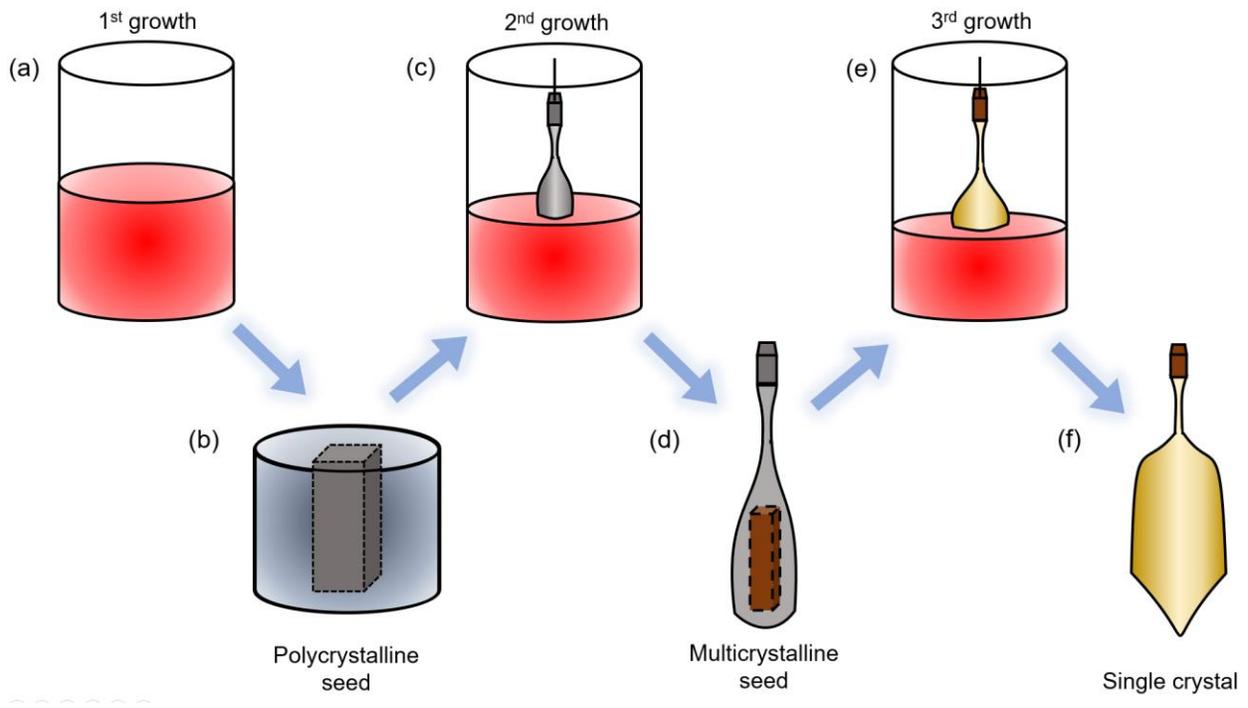

**Figure 1.** Schematic of Czochralski growth of $Cu_3Sn$ single crystal. (a) Cu and Sn granular were melted in the 1$^{st}$ growth. (b) The obtained bulk $Cu_3Sn$ from the 1$^{st}$ growth was cut into a rectangular-shaped seed. (c) In the 2$^{nd}$ growth, the polycrystalline seed was used. (d-f) As-grown multicrystalline $Cu_3Sn$ was cut into the seed again and used for the 3$^{rd}$, 4$^{th}$, and 5$^{th}$ growths.

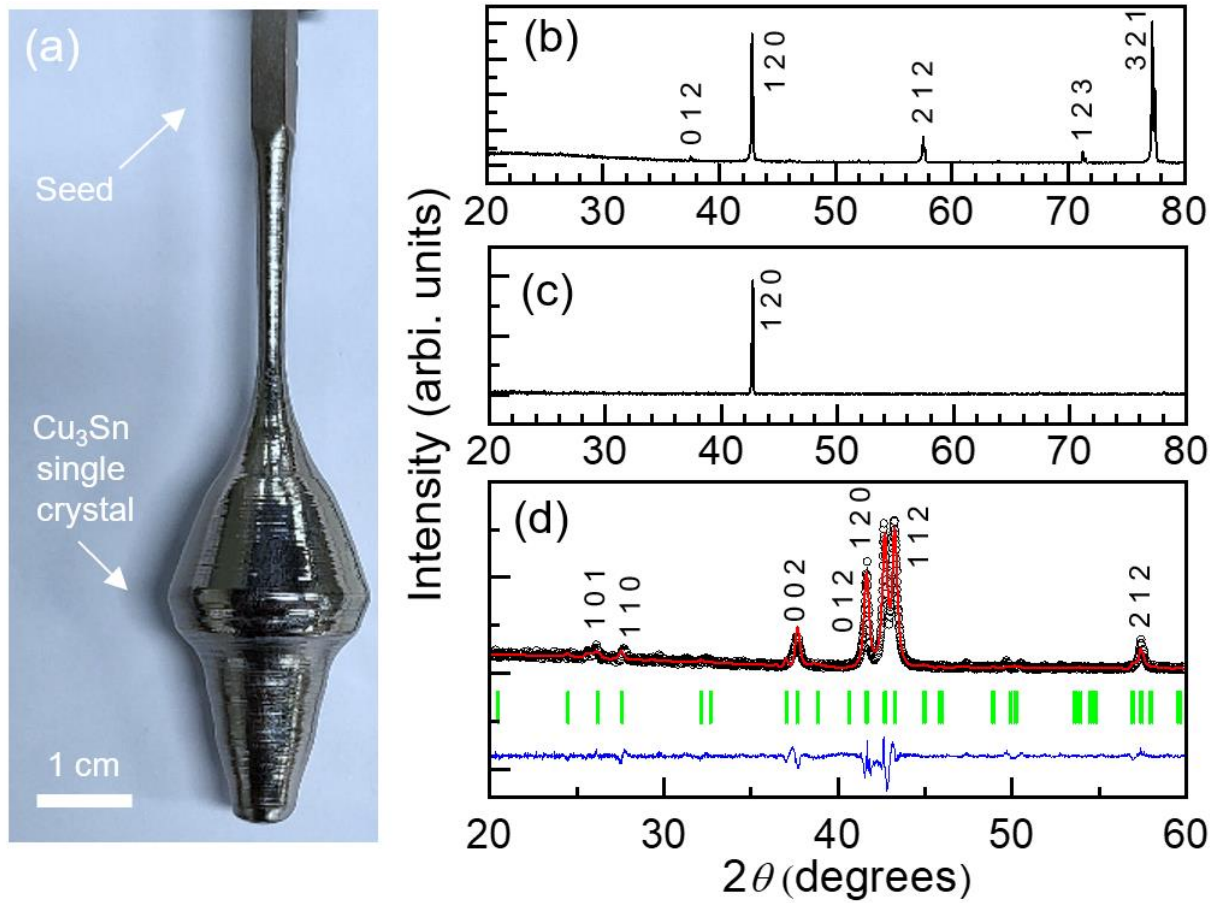

**Figure 2**. (a) The optical image of obtained $Cu_3Sn$ single crystal. (b,c) X-ray diffraction (XRD) pattern of $Cu_3Sn$ crystals from the (b) 2nd growth and (c) 3rd growth. (d) Powder XRD pattern of the ground $Cu_3Sn$ single crystal. Black circles, solid red and blue lines show the observed intensity, the calculated pattern, and the difference between the observed and calculated patterns. Vertical green bars indicate the Bragg peak's positions.

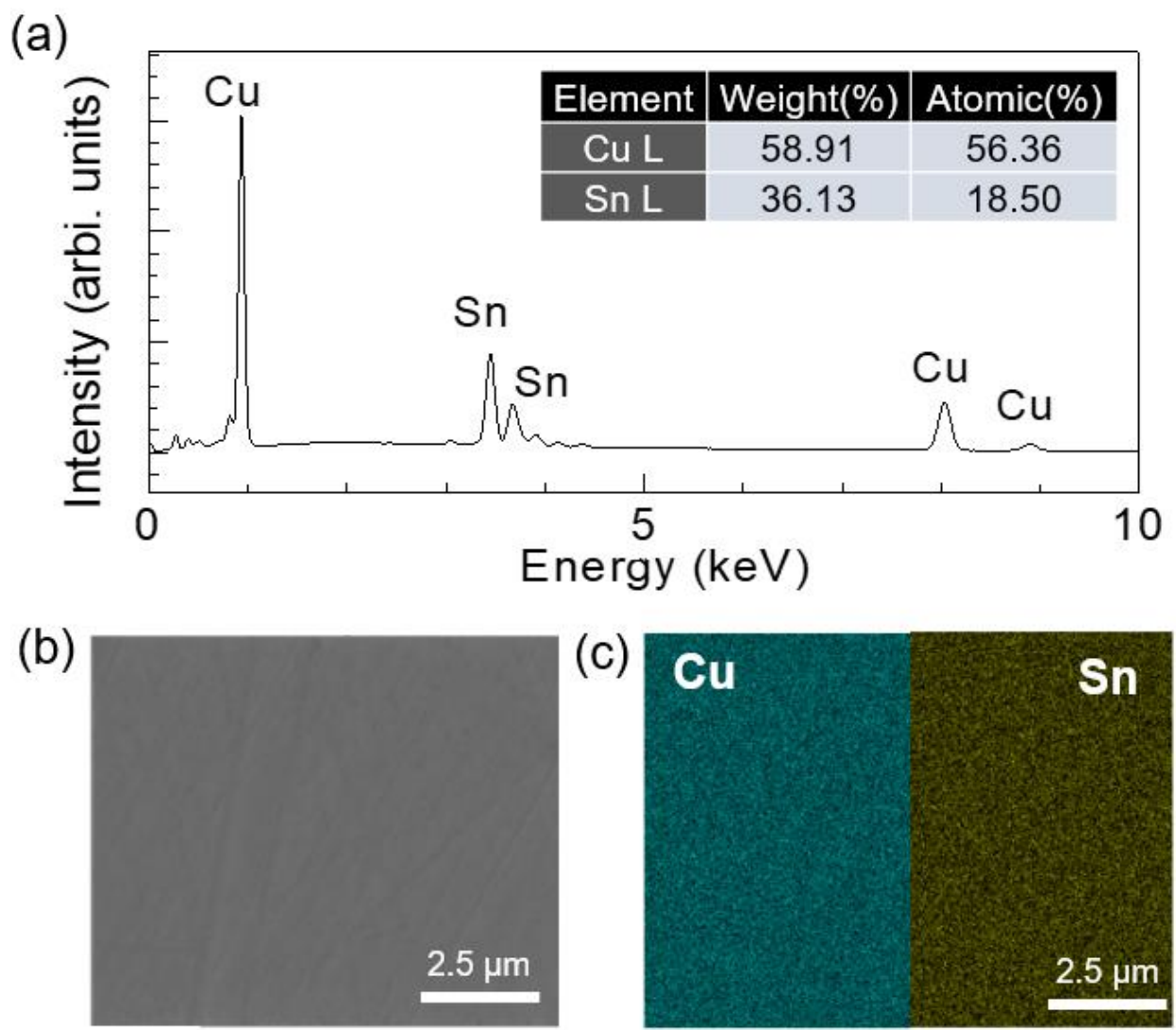

**Figure 3**. (a) Energy-dispersive X-ray (EDS) spectrum of $Cu_3Sn$ single crystal, confirming the existence of Cu and Sn elements. The inset table shows weight and atomic percentage of Cu and Sn. (b) Scanning electron microscope image of $Cu_3Sn$ single crystal. (c) EDS element mapping of $Cu_3Sn$ single crystal, showing the distribution of Cu (left) and Sn (right).

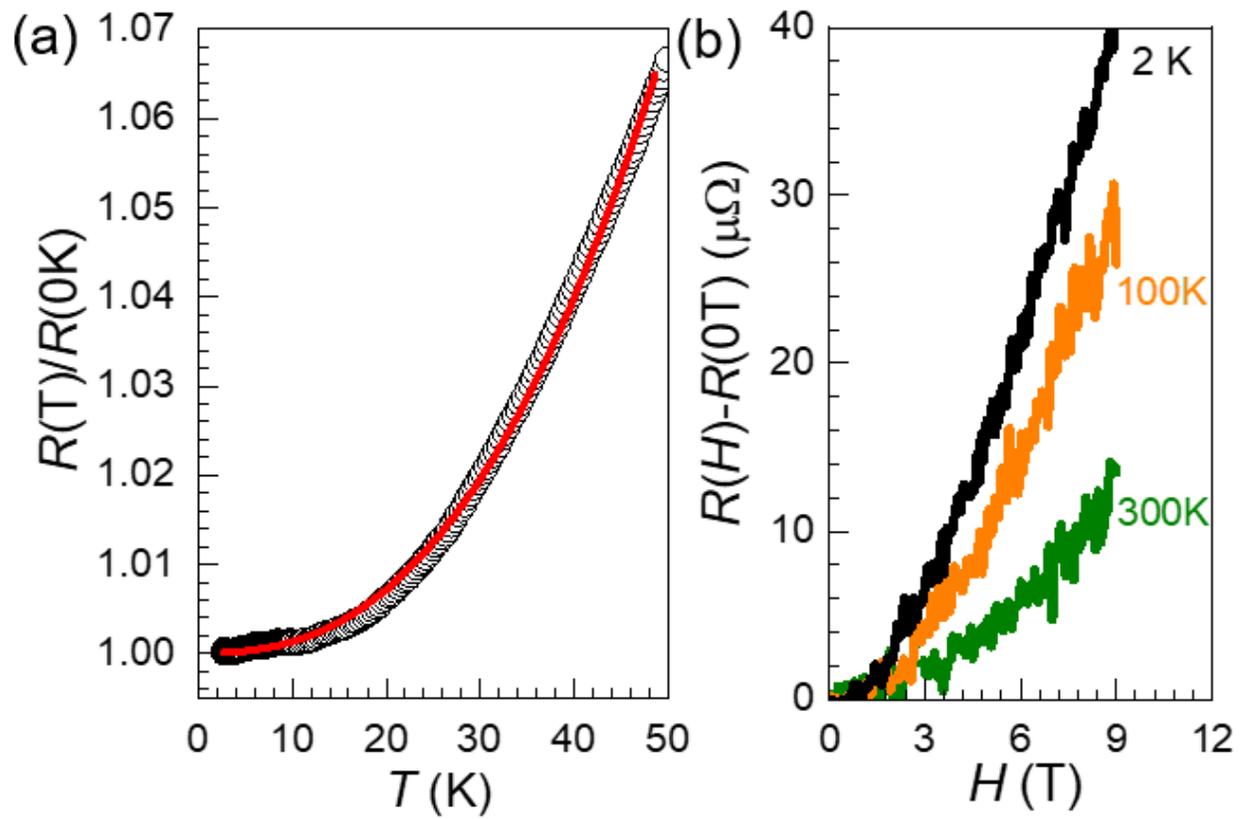

**Figure 4**. (a) Temperature dependence of electrical resistance. The resistance can be fitted using a power-law dependence $T^{\alpha}$ ($\alpha=2.6$) (Red solid line). (b) Resistance changes under magnetic field up to 9 T at various temperatures.

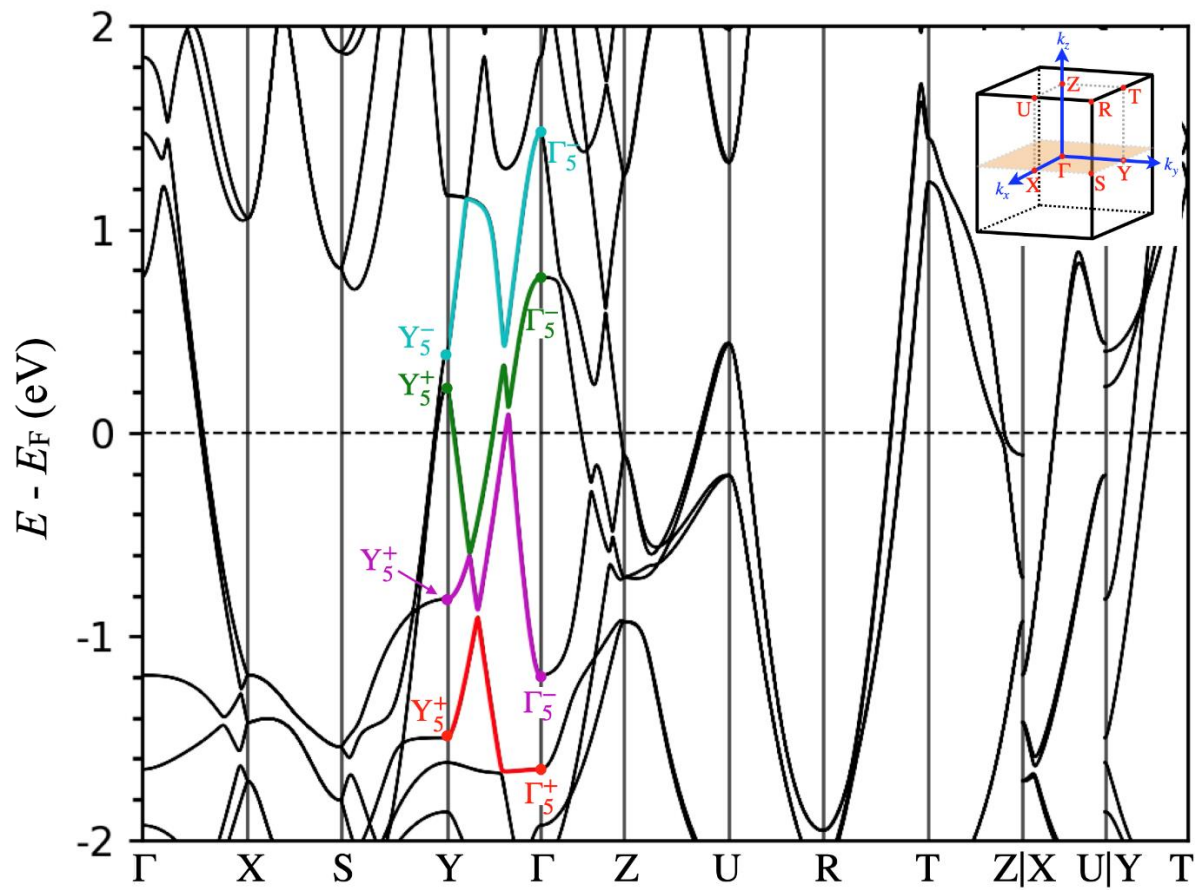

**Figure 5.** Electronic band-structure of orthorhombic *Pmmn* Cu$_3$Sn along with some high symmetry directions. The corresponding Brillouin zone is displayed as an insert.

Table 1. Detailed structural information on $Cu_3Sn$ single crystal obtained from Rietveld refinement parameters. The space group and lattice parameters are *Pmnm*, $a = 4.34$ Å, $b = 4.86$ Å, $c = 5.48$ Å, and $V=115.54$ Å$^3$, respectively.

| Atom position | x | y | z | $B_{iso}$ | Occ |
|---|---|---|---|---|---|
| Cu1 | 0 | 0 | 0.10866 | 6.72100 | 0.27871 |
| Cu2 | 0 | 0 | 0 | 4.50954 | 0.09486 |
| Cu3 | 0 | 0 | 0.49129 | 4.40054 | 0.10902 |
| Cu4 | 0 | 0 | 0 | 0 | 0 |
| Sn1 | 0 | 0 | 0 | 0 | 0 |

Table 2. The products of parity eigenvalues of the occupied states for high symmetry k-points in the bulk Brillouin zone displayed in the inset of Fig. 5. Note that core bands are excluded in the products of parity eigenvalues. Band (parity) inversion occurs at Γ, thereby providing the nontrivial $Z_2$ index.

| Γ | Y | S | X | Z | T | R | U | $Z_2$ |
|---|---|---|---|---|---|---|---|---|
| -1 | +1 | +1 | +1 | +1 | +1 | +1 | +1 | Nontrivial |